\documentstyle[epsfig,12pt]{article}

\def\beqra{\begin{eqnarray}} \def\eeqra{\end{eqnarray}}
\def\beq{\begin{equation}} \def\eeq{\end{equation}}
\def\ds{\displaystyle}

\def \lta {\mathrel{\vcenter
{\hbox{$<$}\nointerlineskip\hbox{$\sim$}}}} \def \gta
{\mathrel{\vcenter {\hbox{$>$}\nointerlineskip\hbox{$\sim$}}}}
\begin{document}
\sloppy
\title{\bf SUSY QCD and Quintessence}
\author{A.  Masiero$^{a,b}$\footnote{e-mail: masiero@sissa.it} ,
M.  Pietroni$^{c}$\footnote{e-mail: pietroni@padova.infn.it} , and
F.  Rosati$^{a,b}$\footnote{e-mail: rosati@sissa.it}} \maketitle
\footnotesize{
\begin{center}
$^a$SISSA, via Beirut 2-4, I-34013 Trieste, Italy\\ $^b$INFN, Sezione
di Trieste, Area di Ricerca, Padriciano 99, I-34012 Trieste, Italy\\ 
$^c$INFN, Sezione di Padova, Via F. Marzolo 8, I-35131 Padova, Italy\\
\end{center}
}

\abstract{Recent data point in the direction of a cosmological constant 
dominated universe. We investigate the r\^ole of supersymmetric QCD with 
$N_f<N_c$ as a possible candidate for dynamical cosmological constant 
(``quintessence'').
When $N_f>1$, the multiscalar dynamics is fully taken into account, 
showing that a certain degree of flavor symmetry in the  initial 
conditions is necessary in order to attain the required late-time 
``tracking'' attractor behavior.
A possible solution to the problem of long-range interactions of gravitational
strength is proposed. Finally we show how, as a consequence of the 
interaction with the Standard Model fields, the early cosmological evolution 
of the scalar fields may be dramatically altered, and the approach to the 
tracking behavior may become much more efficient.}

\vskip0.3in
\noindent
DFPD/99/TH/17\\
SISSA/99/EP/52



%
%
%
\section{Introduction}

Indications for an accelerating universe coming from redshift-distance
measurements of High-Z Supernovae Ia (SNe\ Ia) \cite{scp,highz},
combined with CMB data \cite{cmb} and cluster mass distribution
\cite{cluster}, have recently drawn a great deal of attention on
cosmological models with $\Omega _{m}\sim 1/3$ and $\Omega _{\Lambda
}\sim 2/3$, $\Omega _{m}$ and $\Omega _{\Lambda }$ being the fraction
densities in matter and cosmological constant, respectively.  
More generally, the r\^ole of the cosmological constant in accelerating the
universe expansion could be played by any smooth component with
negative equation of state $p_{Q}/\rho _{Q}=w_{Q}\lta -0.6$
\cite{friem,quint1}, as in the so-called ``quintessence'' models
(QCDM) \cite{quint1}, otherwise known as $x$CDM models \cite{turner1}. 

A natural candidate for quintessence is given by a rolling scalar
field $Q$ with potential $V(Q)$ and equation of state
\[\ds
w_{Q}=\frac{\ds \dot{Q}^{2}/2 -V(Q)}{\ds \dot{Q}^{2}/2 +V(Q)}\;,
\]
which -- depending on the amount of kinetic energy -- could in
principle take any value from $-1$ to $+1$.  The study of scalar field
cosmologies has shown \cite{rp,liddle} that for certain potentials
there exist attractor solutions that can be of the ``scaling''
\cite{wet,cop,fj} or ``tracker'' \cite{zws,swz} type; that means that
for a wide range of initial conditions the scalar field will rapidly
join a well defined late time behavior. 

If $\rho _{Q}\ll \rho _{B}$, where $\rho _{B}$ is the energy
density of the dominant background (radiation or matter), the
attractor can be studied analytically. 

In the case of an exponential potential, $V(Q)\sim \exp {(-Q)}$ the
solution $Q\sim \ln {t}$ is, under very general conditions, a ``scaling''
attractor in phase space characterized by $\rho _{Q}/\rho _{B}\sim
{\rm const}$ \cite{wet,cop,fj}. 
This could potentially solve the so called ``cosmic coincidence'' problem, 
providing a dynamical explanation for the order of magnitude equality 
between matter and scalar field energy today.  
Unfortunately, the equation of state for this attractor is $w_{Q}=w_{B}$, 
which cannot explain the acceleration of the universe neither during RD 
($ w_{rad}=1/3$) nor during MD ($w_m=0$).  
Moreover, Big Bang nucleosynthesis constrain the field
energy density to values much smaller than the required $ \sim 2/3$
\cite{liddle,cop,fj}. 

If instead an inverse power-law potential is considered,
$V(Q)=M^{4+\alpha }Q^{-\alpha }$, with $\alpha >0$, the attractor
solution is $Q\sim t^{1-n/m}$, where $n=3(w_{Q}+1)$, $m=3(w_{B}+1)$;
and the equation of state turns out to be $w_{Q}=(w_{B}
\,\alpha-2)/(\alpha+2)$, which is always negative during MD. 
The ratio of the energies is no longer constant but scales as\ 
$\rho _{Q}/\rho_{B}\sim a^{m-n}$ thus growing during the cosmological 
evolution, since $n$ $<m$.  
$\rho _{Q}$ could then have been safely small during
nucleosynthesis and have grown lately up to the phenomenologically
interesting values.These solutions are then good candidates for
quintessence and have been denominated ``tracker'' in the literature
\cite{liddle,zws,swz}. 

The inverse power-law potential does not improve the cosmic
coincidence problem with respect to the cosmological constant
case. Indeed, the scale $M $ has to be fixed from the requirement that
the scalar energy density today is exactly what is needed. This
corresponds to choosing the desired tracker path.  An important
difference exists in this case though.  
The initial conditions for the physical variable $\rho _{Q}$ can vary 
between the present critical energy density $\rho_{cr}^0$ and the 
background energy density $\rho_B$ at the time of beginning \cite{swz} 
(this range can span many tens of orders of magnitude, depending on the 
initial time), and will anyway end on the tracker path before the present 
epoch, due to the presence of an attractor in phase space \cite{zws,swz}.  
On the contrary, in the cosmological constant case, the physical variable
$\rho _{\Lambda }$ is fixed once for all at the beginning. This allows
us to say that in the quintessence case the fine-tuning issue, even if
still far from solved, is at least weakened. 

A great effort has recently been devoted to find ways to constrain
such models with present and future cosmological data in order to
distinguish quintessence from $\Lambda $ models \cite{constr,det}.  
An even more ambitious goal is the partial reconstruction of the scalar field
potential from measuring the variation of the equation of state with
increasing redshift \cite{turner2}. 

On the other hand, the investigation of quintessence models from the
particle physics point of view is just in a preliminar stage and a
realistic model is still missing (see for example refs. 
\cite{bin,pngb,lyth,sugra}). 
There are two classes of problems; the construction of a field theory model 
with the required scalar potential and the interaction of the quintessence 
field with the standard model (SM) fields \cite{car}.  
The former problem was already considered by Bin\'{e}truy \cite{bin}, 
who pointed out that scalar inverse power law potentials appear in 
supersymmetric QCD theories (SQCD) \cite{SQCD} with $N_{c}$ colors 
and $N_{f}<N_{c}$ flavors. 
The latter seems the toughest. Indeed the quintessence field today has 
typically a mass of order $H_{0}\sim 10^{-33}$eV. 
Then, in general, it would mediate long range interactions of gravitational 
strength, which are phenomenologically unacceptable. 

In this paper we will consider in more detail these problems in the
framework of SQCD. In our analysis we develope two aspects. First of
all, these models have $N_{f}$ independent scalar directions in the
vacuum manifold.  
In ref. \cite{bin} all of them were given the same
initial conditions, so that the dynamics reduced effectively to that
of a single scalar field with an inverse power law potential of the
type considered above.  
On the other hand, in a cosmological setting
there is no a priori justification for this assumption, and the fields
will in general start from different initial conditions. The dynamics
is then truly $N_f$-dimensional and it is relevant to know what is the
late time behavior of the system, whether there are other attractors
besides the single-scalar one discussed in ref. \cite{bin}, and under
what conditions is the latter reached by the system. 
Indeed, we will see that starting with the same initial total energy but 
different initial conditions for the $N_f$ fields may prevent them to 
reach the attractor, so that SQCD cannot be considered as a simple 
one-scalar model for quintessence in these regions of initial conditions 
phase space. 

The second aspect that we consider is the interaction of the SUSY
$SU(N_c)$ model with the SM (or supersymmetric SM (SSM)) fields. The
quintessence fields are usually assumed to be singlets under the SM
gauge group and to interact with the rest of the world only
gravitationally, {\it i.e.}  via non renormalizable operators
suppressed by inverse powers of the Planck mass. 
This is however not enough. In order to prevent long-range interactions 
of gravitational strength it is necessary to assume that the - a priori 
unknown - couplings between the quintessence fields and the SM sector are
strongly suppressed today. 
We do not solve this problem, but point out that if a {\it least coupling 
principle} of the type proposed by Damour and Polyakov \cite{dam3} for the 
superstring dilaton were operative, quintessence models could be reconciled 
with the experimental constraints on the weak equivalence principle and on time
variation of the SM coupling constants. 
At the same time, during RD it would be quite likely to have SUSY breaking 
and mass generation for the quintessence fields, with masses proportional 
to $H$, by the same mechanism discussed by Dine, Randall, and Thomas in 
\cite{drt}.  
If present, these time-dependent SUSY breaking masses would prevent the
fields to take large values, thus driving the system towards the
tracker solution. 

The paper is organized as follows. In Section 2 we review SQCD theories
with $N_f<N_c$ and derive the equation of motion of the scalar degrees
of freedom in the cosmological background. In Section  3 the tracker
solution is discussed.  In Section 4 we address the problem of the
interaction of the superlight fields with the other SM fields and
propose a possible line of solution. In Section 5 numerical results are
presented, with and without the interaction with other
fields. Finally, in Section 6, we present our conclusions. 

\section{SUSY QCD}

As already noted by Bin\`{e}truy \cite{bin}, supersymmetric QCD
theories with $N_{c}$ colors and $N_{f}<N_{c}$ flavors \cite{SQCD} may
give an explicit realization of a model for quintessence with an
inverse power law scalar potential. 
The remarkable feature of these theories is that the superpotential is 
exactly known non-perturbatively. Moreover, in the range of field values 
that will be relevant for our purposes (see below) quantum corrections to the
K\"{a}hler potential are under control. 
As a consequence, we can study the scalar potential and the field equations 
of motion of the full quantum theory, without limiting ourselves to the 
classical approximation. 

The matter content of the theory is given by the chiral superfields
$Q_{i}$ and $\overline{Q}_{i}$ ($i=1\ldots N_{f}$) transforming
according to the $ N_{c}$ and $\overline{N}_{c}$ representations of
$SU(N_c)$, respectively.  In the following, the same symbols will be
used for the superfields $Q_{i}$, $\overline{Q}_{i}$, and their scalar
components. 

Supersymmetry and anomaly-free global symmetries constrain the
superpotential to the unique {\it exact} form
\begin{equation}
W=(N_{c}-N_{f})\left( \frac{\Lambda ^{(3N_{c}-N_{f})}}{{\rm
det}T}\right) ^{ \frac{1}{N_{c}-N_{f}}} \label{superpot}
\end{equation}
where the gauge-invariant matrix superfield $T_{ij}=Q_{i}\cdot
\overline{Q}_{j}$ appears. $\Lambda $ is the only mass scale of the
theory.  
It is the supersymmetric analogue of $\Lambda _{QCD}$, the
renormalization group invariant scale at which the gauge coupling of
$SU(N_{c})$ becomes non-perturbative. As long as scalar field values
$Q_{i},\overline{Q}_{i}\gg $ $\Lambda $ are considered, the theory is
in the weak coupling regime and the canonical form for the K\"{a}hler
potential may be assumed.  
The scalar and fermion matter fields have then canonical kinetic terms, 
and the scalar potential is given by
\begin{equation}
V(Q_{i},\overline{Q}_{i})=\sum_{i=1}^{N_{f}}\left(
|F_{Q_{i}}|^{2}+|F_{\overline{Q}_{i}}|^{2}\right)
+\frac{1}{2}D^{a}D^{a} \label{potscal}
\end{equation}
\label{scalarpot} 
where $F_{Q_{i}}=\partial W/\partial Q_{i}$, 
$F_{\overline{Q}_{i}}=\partial W/\partial \overline{Q}_{i}$, and
\begin{equation}
D^{a}=Q_{i}^{\dagger }t^{a}Q_{i}-\overline{Q}_{i}t^{a}\overline{Q}
_{i}^{\dagger }\;.  \label{d-terms}
\end{equation}
The relevant dynamics of the field expectation values takes place
along directions in field space in which the above D-term vanish, {\it
i.e.} the perturbatively flat directions $\langle Q_{i\alpha }\rangle
=\langle \overline{Q}_{i\alpha }^{\dagger }\rangle $, where $\alpha
=1\cdots N_{c}$ is the gauge index.  
At the non-perturbative level these directions get a non vanishing 
potential from the F-terms in (\ref{potscal}), which are zero at any order 
in perturbation theory.  
Gauge and flavor rotations can be used to diagonalize the
$\langle Q_{i\alpha }\rangle $ and put them in the form
\[
\langle Q_{i\alpha }\rangle =\langle \overline{Q}_{i\alpha }^{\dagger
}\rangle =
\begin{array}{l}
q_{i}\delta _{i\alpha }\;\;\;\;1\leq \alpha \leq N_{f} \\ 0\;\;\;\;\ \
\ \ \;N_{f}\leq \alpha \leq N_{c}
\end{array}
. 
\]
Along these directions, the scalar potential is given by
\begin{eqnarray*}
v(q_{i}) &\equiv &\langle V(Q_{i},\overline{Q}_{i})\rangle
=2\;\frac{\Lambda ^{2a }}{\prod_{i=1}^{N_{f}}|q_{i}|^{4d
}}\;\left( \sum_{j=1}^{N_{f}}\frac{1}{|q_{j}|^{2}}\right) ,\qquad \\
&& \\ \qquad a &=&\frac{3N_{c}-N_{f}}{N_{c}-N_{f}},\ \ \ \ \ \
d =\frac{1}{N_{c}-N_{f}}. 
\end{eqnarray*}
In the following, we will be interested in the cosmological evolution
of the $N_{f}$ expectation values $q_{i}$, given by
\[
\langle \ddot{Q_{i}}+3H\dot{Q_{i}}+\frac{\partial V}{\partial
Q_{i}^{\dagger }}\rangle =0\;\;,\;i=1,...,N_{f}\;. 
\]
In ref.  \cite{bin} the same initial conditions for all the $N_{f}$
VEV's and their time derivatives were chosen. With this very peculiar
choice the evolution of the system may be described by a single VEV
$q$ (which we take real) with equation of motion 
\beq
\ddot{q}+3H\dot{q}-g \frac{\Lambda ^{2a }}{q^{2g
+1}}=0\ ,\qquad \qquad g =\frac{N_{c}+N_{f}}{N_{c}-N_{f}}\ ,
\label{onescalar}
\eeq 
thus reproducing exactly the case of a single scalar field $\Phi
$ in the potential $V=\Lambda ^{4+2g}\Phi ^{-2g}/2$
considered in refs.  \cite{rp,liddle,swz}.  In this paper we will
consider the more general case in which different initial conditions
are assigned to different VEV's, and the system is described by
$N_{f}$ coupled differential equations. Taking for illustration the
case $N_{f}=2$, we will have to solve the equations

\[
\ddot{q}_{1}+3H\dot{q}_{1}-d\cdot\! \ q_{1}\ \frac{\Lambda ^{2a
}}{\left( q_{1}q_{2}\right) ^{2d N_{c}}}\ \left[
2+N_{c}\frac{q_{2}^{2}}{q_{1}^{2}}\right] =0\ ,
\]
 
\begin{equation}
\ddot{q}_{2}+3H\dot{q}_{2}-d\cdot\! \ q_{2}\ \frac{\Lambda ^{2a
}}{\left( q_{1}q_{2}\right) ^{2d N_{c}}}\ \left[
2+N_{c}\frac{q_{1}^{2}}{q_{2}^{2}}\right] =0\ , \label{eom}
\end{equation}
with $H^{2}=8\pi /3M_P^{2}\ (\rho _{m}+\rho _{r}+\rho _{Q})$, where
$M_P$ is the Planck mass, $\rho _{m(r)}$ is the matter (radiation)
energy density, and $\rho
_{Q}=2(\dot{q}_{1}^{2}+\dot{q}_{2}^{2})+v(q_{1},q_{2})$ is the total
field energy. 

\section{The tracker solution}

In analogy with the one-scalar case, we look for power-law{\it \
}solutions of the form
\begin{equation}
q_{tr,i}=C_{i}\cdot t^{\, p_{i}}\ ,\qquad \qquad i=1,\cdots ,\ N_{f}\ . 
\label{scaling}
\end{equation}
It is straightforward to verify that -- when $\rho _{Q}\ll \rho _{B}$
-- the only solution of this type is given by
\[
p_{i}\equiv p=\frac{1-r}{2}\ ,\ \ \ \ \ \ C_{i}\equiv C=\left[
X^{1-r}\ \Lambda ^{2(3-r)}\right] ^{1/4}\ ,\ \ \
\ \ \ i=1,\cdots ,\ N_{f}\ ,
\]
with
\[
X\equiv \frac{4\ m\ \ (1+r)}{(1-r)^{2}\ [12-m(1+r)]}\ ,
\]
where we have defined $r\equiv N_{f}/N_c$ $(=1/N_{c},\ldots ,1-1/N_{c})$.  
This solution is characterized by an equation of state
\begin{equation}
w_{Q}=\frac{1+r}{2}w_{B}-\frac{1-r}{2}\ .  \label{eosfree}
\end{equation}
Eq. (\ref{eosfree}) can be derived as usual from energy conservation
{\it i.e. }$d\ (a^3 \rho_{Q})=-3\ a^2 p_{Q}$. 

Following the same methods employed in ref. \cite{liddle} one can show
that the above solution is the unique stable attractor in the space of
solutions of eqs. (\ref{eom}). Then, even if the $q_{i}$'s start with
different initial conditions, there is a region in field configuration
space such that the system evolves towards the equal fields solutions
(\ref{scaling}), and the late-time behavior is indistinguishable from
the case considered in ref.  \cite{bin}. 

The field energy density grows with respect to the matter energy
density as
\begin{equation}
\frac{\rho _{Q}}{\rho _{m}}\sim a^{\frac{3(1+r)}{2}},
\end{equation}
where $a$ is the scale factor of the universe. The scalar field energy
will then eventually dominate and the approximations leading to the
scaling solution (\ref{scaling}) will drop, so that a numerical
treatment of the field equations is mandatory in order to describe the
phenomenologically relevant late-time behavior. 

The scale $\Lambda $ can be fixed requiring that the scalar fields are
starting to dominate the energy density of the universe today and that
both have already reached the tracking behavior.  The two conditions
are realized if
\begin{equation}
v(q_{0})\simeq \rho _{crit}^{0}\ ,\qquad \qquad v^{\prime \prime
}(q_{0})\simeq H_{0}^{2}\ , \label{conditions}
\end{equation}
where $\rho _{crit}^{0}=3M_P^{2}H_{0}^{2}/8\pi $ and $q_{0}$ are 
the present critical density and scalar fields VEV respectively. 
Eqs. (\ref{conditions}) imply
\begin{eqnarray}
\frac{\Lambda }{M_P} & \simeq &\left[ \frac{3}{4\pi
}\frac{(1+r)(3+r)}{(1-r)^{2}} \frac{1}{rN_c}\right] ^{\frac{1+r}{2(3-r)}}
\left( \frac{1}{2rN_{c}}\frac{\rho _{crit}^{0}}{M_P^{4}}\right)
^{\frac{1-r}{2(3-r)}}\ , \label{today} \\ && \nonumber \\
\frac{q_{0}^{2}}{M_P^{2}} & \simeq &\frac{3}{4\pi
}\frac{(1+r)(3+r)}{(1-r)^{2}} \frac{1}{rN_c}\ . \label{todayvev}
\end{eqnarray} 

Depending on the values for $N_{f}$ and $N_{c}$, $\Lambda $ and $
q_{0}/\Lambda $ assume widely different values. $\Lambda $ takes its
lowest possible values in the $N_{c}\rightarrow \infty $ ($N_{f}$
fixed) limit, where it equals \mbox{$4\cdot
10^{-2}(h^{2}/N_{f}^2)^{1/6}\ $GeV} (we have used $\rho
_{crit}^{0}/M_P^{4}=(2.5\cdot 10^{-31}h^{1/2})^{4}$).  For fixed
$N_{c}$, instead, $\Lambda $ increases as $N_{f}$ goes from $1$ to its
maximum allowed value, $N_{f}=1-N_{c}$.  For $N_{c}\gta 20$ and
$N_{f}$ close to $N_{c}$, the scale $\Lambda $ exceeds $M_P.$

The accuracy of the determination of $\Lambda $ given in (\ref{today})
depends on the present error on the measurements of $H_{0}$, {\it
i.e., } typically,{\it \ }$\delta \Lambda /\Lambda =\frac{1-r}{3-r}\ 
\delta H_{0}/H_{0}\lta 0.1$. 

In deriving the scalar potential (\ref{potscal}) and the field
equations (\ref{eom}) we have assumed that the system is in the
weakly coupled regime, so that the canonical form for the K\"{a}hler
potential may be considered as a good approximation. This condition is
satisfied as long as the fields' VEVs are much larger than the
non-perturbative scale $\Lambda $.  From eqs. (\ref{today}) and
(\ref{todayvev}), one can compute the ratio between the VEVs today and
$\Lambda $, and see that it is greater than unity for any $N_f$ as
long as $N_c \lta 20$. 

On the other hand, if we want to follow the cosmological evolution of
the fields starting from an earlier cosmological epoch, we must impose
the stronger condition that the $q_{i}$ have been much larger than
$\Lambda $ throughout the time interval of interest.  Taking the
tracker solution (\ref {scaling}) as a reference, at the initial
redshift $z_{in}$ we have,
\[
\frac{q_{tr}^{in}}{\Lambda }=\frac{q_{tr,0}}{\Lambda
}\frac{q_{tr}^{in}}{q_{tr,0}}=\frac{q_{tr,0}}{\Lambda }~\langle
\begin{array}{l}
z_{in}^{r-1}\ z_{eq}^{(1-r)/4}\qquad \qquad {\mathrm if~~~
}z_{in}>z_{eq} \\ \\ z_{in}^{3(r-1)/4}\qquad ~~~~~~\qquad {\mathrm if
~~~}z_{in}\leq z_{eq}
\end{array}
,
\]
where $z_{eq}$ is the redshift at matter-radiation equivalence. For a
given $r$, the condition $q_{tr}^{in}\gg \Lambda $ gives an upper bound
on $z_{in}$.  Taking for instance $N_{c}\rightarrow \infty $ ($N_{f}$
fixed) we get $z_{in}\ll 10^{21}(N_{f}h)^{-1/3}$. In the numerical
computations, we will consider initial conditions such that the weak
coupling regime is always realized. 

\section{Interaction with the visible sector}

The superfields $Q_{i}$ and $\overline{Q}_{i}$ have been taken as
singlets under the SM gauge group. Therefore, they may interact with
the visible sector only gravitationally, {\it i.e.  }via
non-renormalizable operators suppressed by inverse powers of the
Planck mass, of the form
\begin{equation}
\int d^{4}\theta \ K^{j}(\phi _{j}^{\dagger },\phi _{j})\ \left[ \beta
_{n}^{ji}\frac{\left( Q_{i}^{\dagger }Q_{i}\right) ^{n}}{M_P^{2n}}+
\overline{\beta }_{n}^{ji}\frac{\left( \overline{Q}_{i}\overline{Q}
_{i}^{\dagger }\right) ^{n}}{M_P^{2n}}\right] \ \ , \label{coupling}
\end{equation}
where $\phi _{j}$ represents a generic standard model superfield. From
(\ref {today}) we know that today the VEV's $q_{i}$ are typically
$O(M_P)$, so there is no reason to limit ourselves to the
contributions of lowest order in $|Q|^{2}/M_P^{2}$. Rather, we have
to consider the full (unknown) functions
\[
\beta ^{ji}\left[ \frac{Q_{i}^{\dagger }Q_{i}}{M_P^{2}}\right] \
\equiv \sum_{n=0}^{\infty }\beta _{n}^{ji}\frac{\left( Q_{i}^{\dagger
}Q_{i}\right) ^{n}}{M_P^{2n}},\ \
\]
and the analogous $\overline{\beta }$'s for the $\overline{Q}_{i}$'s. 
Moreover, the requirement that the scalar fields are on the tracking
solution today, eqs. (\ref{conditions}) implies that their mass is of
order $\sim H_{0}^{2}\sim 10^{-33}$ eV. 

The exchange of very light fields gives rise to long-range forces
which are constrained by tests on the equivalence principle, whereas
the time dependence of the VEV's induces a time variation of the SM
coupling constants \cite{car,dam}.  These kind of considerations sets
stringent bounds on the first derivatives of the $\beta ^{ji}$'s and
$\overline{\beta }^{ji}$'s {\it today,}

\[
\alpha ^{ji}\equiv \left.\frac{d\log \beta ^{ji}\left[ x_i^2 \right]
}{d x_i}\right|_{x_i=x_i^0} ,\qquad \qquad \overline{\alpha
}^{ji}\equiv \left.\frac{d\log \overline{\beta }^{ji}\left[ x_i^2
\right] }{d\,x_i}\right|_{x_i=x_i^0} \qquad,
\]
where $x_i \equiv q_i/M_P$.  To give an example, the best bound on the
time variation of the fine structure constant comes from the Oklo
natural reactor. It implies that $\left| \dot{\alpha}/\alpha \right|
<10^{-15}\ {\rm yr}^{-1}$ \cite{dam2}, leading to the following
constraint on the coupling with the kinetic terms of the
electromagnetic vector superfield $V$,
\begin{equation}
\alpha ^{Vi},\ \overline{\alpha }^{Vi}\ \lta\ 10^{-6}\ \frac{H_{0}}{
\left\langle \dot{q}_{i}\right\rangle }\,M_P \,, \label{decoupling}
\end{equation}
where $\left\langle \dot{q}_{i}\right\rangle $ is the average rate of
change of $q_{i}$ in the past $2\times 10^{9}{\rm yr}$. 

Similar --although generally less stringent-- bounds can be
analogously obtained for the coupling with the other standard model
superfields \cite{dam}. Therefore, in order to be phenomenologically
viable, any quintessence model should postulate that all the unknown
couplings $\beta ^{ji}$'s and $\overline{\beta }^{ji}$'s have a
common minimum close to the actual value of the $q_{i}$'s\footnote{
An alternative way to suppress long-range interactions, based on an
approximate global symmetry, was proposed in ref. \cite{car}.}. 

The simplest way to realize this condition would be via the {\it least
coupling principle } introduced by Damour and Polyakov for the
massless superstring dilaton in ref. \cite{dam3}, where a universal
coupling between the dilaton and the SM fields was postulated. In the
present context, we will invoke a similar principle, by postulating
that $\beta ^{ji}=\beta $ and $\overline{\beta }^{ji}=\overline{\beta
}$ for any SM field $\phi _{j}$ and any flavor $i$. For simplicity, we
will further assume $\beta =\overline{\beta }$ . 

The decoupling from the visible sector implied by bounds like (\ref
{decoupling}) does not necessarily mean that the interactions between
the quintessence sector and the visible one have always been
phenomenologically irrelevant. Indeed, during radiation domination the
VEVs $q_{i}$ were typically $\ll M_P$ and then very far from the
postulated minimum of the $\beta $'s. For such values of the
$q_{i}$'s the $\beta $'s can be approximated as
\begin{equation}
\beta \left[ \frac{Q^{\dagger }Q}{M_P^{2}}\right] =\beta _{0}+\beta
_{1}\frac{Q^{\dagger }Q}{M_P^{2}}\ +\ldots \label{betarad}
\end{equation}
where the constants $\beta _{0}$ and $\beta _{1}$ are not directly
constrained by (\ref{decoupling}).  The coupling between the
(\ref{betarad}) and the SM\ kinetic terms, as in (\ref{coupling}),
induces a SUSY breaking mass term for the scalars of the form
\cite{drt}
\begin{equation}
\Delta L\sim H^{2}\,\beta_1 \sum_{i}\ (\left| Q_{i}\right| ^{2}+\left|
\overline{Q}_{i}\right| ^{2})\ ,\; \label{masses}
\end{equation}
where we have used the fact that during radiation domination
$\left\langle \sum_{j}\int d^{4}\theta K^{j}(\phi _{j}^{\dagger },\phi
_{j})\right\rangle \sim \rho _{rad}$. 

If present, this term would have a very interesting impact on the
cosmological evolution of the fields. First of all one should notice
that, unlike the usual mass terms with time-independent masses
considered in \cite{lyth}, a mass $m^{2}\sim H^{2}$ does not modify
the time-dependence of the tracking solution, which is still the
power-law given in eq. (\ref{scaling}).  Thus, the fine-tuning
problems induced by the requirement that a soft-supersymmetry breaking
mass does not spoil the tracking solutions \cite{lyth} are not present
here. 

Secondly, since the $Q$ and $\overline{Q}$ fields do not form an
isolated system any more, the equation of state of the quintessence
fields is not linked to the power-law dependence of the tracking
solution.  Taking into account the interaction with the SM fields,
represented by $H^{2}$, we find the new equation of state during
radiation domination ($w_{B}=1/3$)

\[
w_{Q}^{\prime }=w_{Q}\,-\,4\beta_1 \,\frac{1+r}{9(1-r)+6\beta_1 }
\]
where $w_{Q}$ was given in eq. (\ref{eosfree}). 

From a phenomenological point of view, the most relevant effect of the
presence of mass terms like (\ref{masses}) during radiation domination
resides in the fact that they rise the scalar potential at large
fields values, inducing a (time-dependent) minimum. In absence of such
terms, if the fields are initially very far from the origin, they are
not able to catch up with the tracking behavior before the present
epoch, and $\rho _{Q}$ always remains much smaller than $\rho
_{B}$.  These are the well-known `undershoot' solutions considered in
ref. \cite{swz}. Instead, when large enough masses (\ref{masses}) are
present, the fields are quickly driven towards the time-dependent
minimum and the would-be undershoot solutions reach the tracking
behavior in time. 

The same happens for the would-be `overshoot' solutions, \cite{swz},
in which the fields are initially very close to the origin, with an
energy density much larger than the tracker one, and are subsequently
pushed to very large values, from where they will not be able to reach
the tracking solution before the present epoch. Introducing mass terms
like (\ref{masses}) prevents the fields to go to very large values,
and keeps them closer to the traking solution. 

In other words, the already large region in initial condition phase
space leading to late-time tracking behavior, will be enlarged to the
full phase space. 
In the next section we will discuss numerical results with and without
the supersymmetry breaking mass (\ref{masses}). 

\section{Numerical results}

In this section we illustrate the general results of the previous
sections for the particular case $N_f=2$,
$N_c=6$.

\begin{figure}[hb]
\begin{center}
\epsfig{file=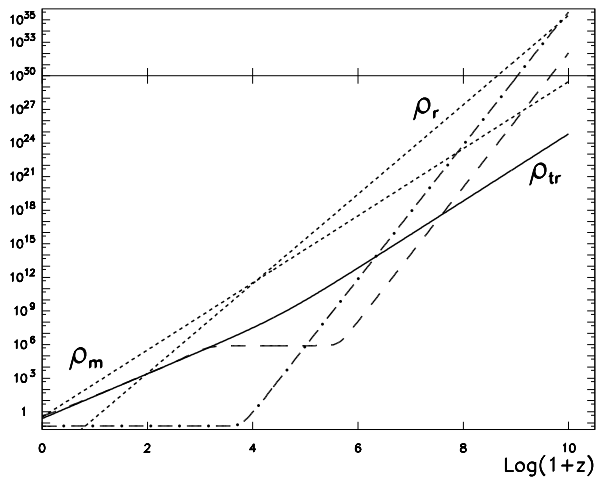,bbllx=30,bblly=200,bburx=560,bbury=600,height=6.3cm}
\begin{minipage}{14cm}
\begin{center}
\caption{\scriptsize \sl  The evolution of
the energy densities $\rho$ of different cosmological components is given 
as a funcion of red-shift. All the energy densities are normalized to the 
present critical energy density $\rho_{cr}^0$. Radiation and matter
energy densities are represented by the short-dashed lines, whereas the
solid line is the energy density of the tracker solution discussed in
Section 3. The long-dashed line is the evolution of the scalar field
energy density for a solution that reaches the tracker before the present 
epoch; while the dash-dotted line represents the evolution for a solution 
that overshoots the tracker to such an extent that it has not yet had enough 
time to re-join the attractor.}
\end{center}
\end{minipage}
\end{center}
\end{figure}

In Fig.1 the energy densities {\em vs.}  redshift are given. We have
chosen the same initial conditions for the two VEVs, in order to
effectively reproduce the one-scalar case of eq. (\ref{onescalar}),
already studied in refs.  \cite{rp,liddle,swz}. 
No interaction with other fields of the type discussed in the
previous section has been considered. 

We see that, depending on the initial energy density of the scalar
fields, the tracker solution may (long dashed line) or may not 
(dash-dotted line) be reached before the present epoch. The latter case
corresponds to the overshoot solutions discussed in ref. \cite{swz},
in which the initial scalar field energy is larger than $\rho_B$ and 
the fields are rapidly pushed to very large values. 
The undershoot region, in which the energy density is always
lower than the tracker one, corresponds to 
$\rho_{cr}^0 \leq \rho_Q^{in} \leq \rho_{tr}^{in}$. 
Thus, all together, there are around 35  orders of magnitude in $\rho_Q^{in}$ 
at redshift $z+1 = 10^{10}$ for which the tracker solution is reached 
before today. Cleary, the more we go backwards in time, the larger is the
allowed initial conditions range. 

In Fig.2 we plot the scalar field equation of state, $w_Q$, for the
corresponding solutions of Fig.1.

\begin{figure}[h]
\begin{center}
\epsfig{file=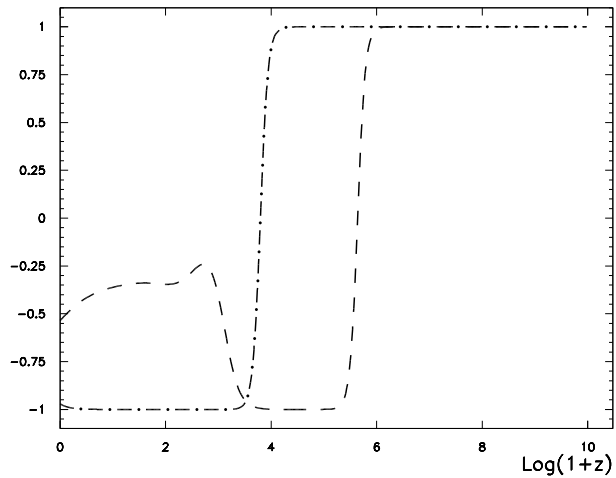,bbllx=30,bblly=200,bburx=560,bbury=600,height=6.3cm}
\begin{minipage}{14cm}
\begin{center}
\caption{\scriptsize \sl The cosmological evolution of the 
equation of state $w_Q \equiv p_Q/ \rho_Q$ for the scalar field $Q$ is 
plotted as a function of red-shift. The two cases correspond to the energy 
densities of the two examples in Fig.1. Note that, in the long-dashed curve
case, the attractor value of $-1/3$ for the equation of state (corresponding 
to $N_c =6)$ is joined well before the present epoch and only very recently 
abandoned, when the scalar field starts to become the dominating component 
of our universe and as a consequence its equation of state is rapidly driven 
towards $-1$.}
\end{center}
\end{minipage}
\end{center}
\end{figure}

Next, we explore to which extent the two-field system that we are
considering behaves as a one scalar model with inverse power-law
potential. In fig. 3 we plot solutions with the same initial energy
density but different ratios between the initial values of the two
scalar fields. 
Given any initial energy density such that -- for $q^{in}_1/q^{in}_2 =1$ -- the
tracker is joined before today, there exists always a limiting value for the
fields' difference above which the attractor is not reached in time.

The effect of the interaction with other fields discussed in Section 4 is
shown in Fig.4. Here, we have included the mass term (\ref{masses})
during radiation domination with $\beta_1 = 0.3$ and we have followed
the same procedure as for Fig.1, searching for undershoot and
overshoot solutions. As we see, the range of initial energy densities
for the solutions reaching the tracker is now enormously enhanced
since, as we discussed previously, the fields are now prevented from
taking too large values. The same conclusion holds even if different
initial conditions for the two fields are allowed, for the same
reason. 

\begin{figure}[ht]
\begin{center}
\epsfig{file=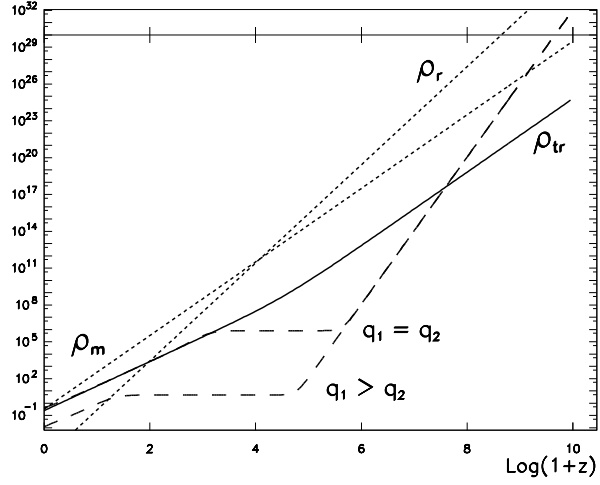,bbllx=30,bblly=200,bburx=560,bbury=600,height=6.3cm}
\begin{minipage}{14cm}
\begin{center}
\caption{\scriptsize \sl The effect of taking different initial conditions
for the fields, at the same initial total field energy.
Starting with $q_1^{in}/q_2^{in} =10^{15}$ 
the tracker behaviour is not reached today. For comparison we plot the
solution for $q_1^{in}/q_2^{in} =1$.}
\end{center}
\end{minipage}
\end{center}
\end{figure}

\vskip0.2cm
\begin{figure}[ht!]
\begin{center}
\epsfig{file=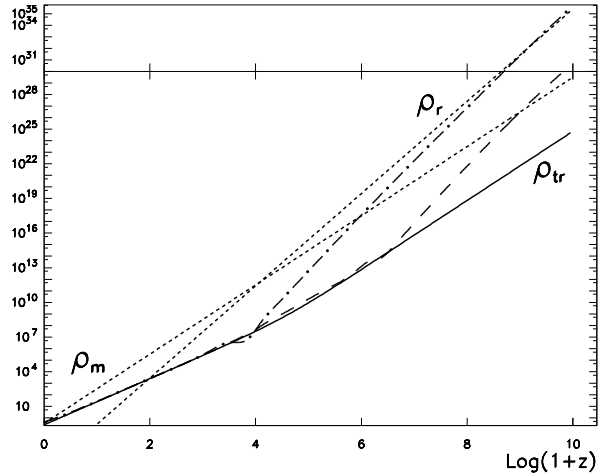,bbllx=30,bblly=200,bburx=560,bbury=600,height=6.3cm}
\begin{minipage}{14cm}
\begin{center}
\caption{\scriptsize \sl The effect of the interaction with other fields, to
be compared with Fig. 1.
Adding a term  like eq. (\ref{masses})
with $\beta_1 = 0.3$ the would-be overshooting solution (dash-dotted line)
reaches the tracker in time.}
\end{center}
\end{minipage}
\end{center}
\end{figure}

\section{Conclusions}

In this paper, we have analyzed the role of SQCD as a possible model
for quintessence.  Our analysis completes the previous one by
Bin\'{e}truy \cite{bin} in two respects. 

First, we have taken in full consideration the multi-scalar nature of
the model, allowing for different initial conditions for the $N_f$
independent scalar VEVs and studying the coupled system of $N_f$
equations of motion.  
Starting with the same initial scalar energy density, but different 
fields' values we have shown that the tracking behavior becomes more 
difficult to reach the larger the difference among the initial conditions 
for the fields. Thus, an approximate flavor symmetry of the initial conditions 
is needed in order that SQCD may act as an effective quintessence model. 

Secondly, we have sketched a possible way out to the common problem
of all quintessence models considered so far, that is the presence of
long-range interactions of gravitational strength mediated by the
ultra-light scalar fields \cite{car}. Our solution is inspired by the
Damour-Polyakov relaxation mechanism for the superstring dilaton
\cite{dam3}. 
Basically, we postulate that all the couplings of the
SQCD quark superfields with the SM ones are given by a unique 
function, which has a minimum close to today's values of the scalar 
fields' VEVs. 
Since all the deviations from Einstein gravity are parametrized by the 
slope of these couplings today, this could make the model phenomenologically
safe with respect to limits on the weak equivalence principle and on
the time dependence of the SM coupling constants. 
At the same time, during radiation domination the coupling with SM 
fields may have induced a SUSY-breaking -time dependent- mass to the 
scalar fields, with the effect of enhancing the initial configuration 
space leading to a late time tracking behavior. 

A deeper investigation of this scenario requires the implementation of 
SQCD in a wider context, such as superstring theories or 
theories with large compact extra-dimensions, what clearly lies beyond 
the scope of the present work.

\end{document}